# Hanle Magnetoresistance in Thin Metal Films with Strong Spin-Orbit Coupling

Saül Vélez[1,*], Vitaly N. Golovach[2,3,4], Amilcar Bedoya-Pinto[1], Miren Isasa[1], Edurne Sagasta[1], Mikel Abadia[2,3], Celia Rogero[2,3], Luis E. Hueso[1,4], F. Sebastian Bergeret[2,3], Fèlix Casanova[1,4,†]

[1]CIC nanoGUNE, 20018 Donostia-San Sebastian, Basque Country, Spain
[2]Centro de  Física de Materiales (CFM-MPC), Centro Mixto CSIC-UPV/EHU, 20018 Donostia-San Sebastian, Basque Country, Spain
[3]Donostia International Physics Center (DIPC), 20018 Donostia-San Sebastian, Basque Country, Spain
[4]IKERBASQUE, Basque Foundation for Science, 48011 Bilbao, Basque Country, Spain

* s.velez@nanogune.eu,

† f.casanova@nanogune.eu

## Abstract

We report measurements of a new type of magnetoresistance in Pt and Ta thin films. The spin accumulation created at the surfaces of the film by the spin Hall effect decreases in a magnetic field because of the Hanle effect, resulting in an increase of the electrical resistance as predicted by Dyakonov [PRL 99, 126601 (2007)]. The angular dependence of this magnetoresistance resembles the recently discovered spin Hall magnetoresistance in Pt/$Y_3Fe_5O_{12}$ bilayers, although the presence of a ferromagnetic insulator is not required. We show that this *Hanle magnetoresistance* is an alternative, simple way to quantitatively study the coupling between charge and spin currents in metals with strong spin-orbit coupling.

Spin-orbit interaction is an essential ingredient in materials and interfaces, offering the possibility to exploit the coupling between spin and orbital degrees of freedom of electrons in spintronic devices [1,2]. Of utmost importance are the spin Hall (SHE) and inverse spin Hall (ISHE) effects, which convert charge currents into transverse spin currents and vice versa, allowing us to create and detect spin currents in materials with strong spin-orbit coupling (SOC) [3–8]. In this framework, a new type of magnetoresistance (MR), spin Hall magnetoresistance (SMR), was discovered in nonmagnetic (NM) metal/ferromagnetic insulator (FMI) bilayers [9–16]. SMR arises from the simultaneous effect of SHE and ISHE in the NM layer –which leads to a decrease in its resistance– combined with the presence of a FMI at one of the interfaces. The spin current generated *via* SHE in the NM layer –when applying a charge current– can be either reflected at the NM/FMI interface when the magnetization **M** of the FMI is parallel to the spin polarization **s** of the spin current in the NM layer or absorbed by the FMI via spin-transfer torque when **M** is perpendicular to **s**, leading to an increase in the resistance of the NM layer in the latter case. As a result, the resistance of the NM layer can be modulated by controlling **M** of the FMI. Although this effect has been mostly studied in NM/FMI bilayers, recent experiments show that SMR is also present when the ferromagnetic (FM) layer is metallic, leading in this case to an enhanced SMR [17–19] and to the emergence of a unidirectional spin Hall magnetoresistance [17]. Despite all this, there is still an intense debate about the physical origin of SMR. For instance, the magnetic proximity effect (MPE) induced in the NM layer in contact with the FM substrate has also been invoked [20–25].

In this Letter, we report a novel MR effect occurring in NM thin films with strong SOC. Our MR measurements are consistent with an effect originally predicted by Dyakonov [26]. The spin accumulation created at the edges of the film by the SHE –in our geometry, the top and bottom surfaces– is suppressed in a magnetic field **H** due to the Hanle effect (spin dephasing arising from simultaneous precession and diffusion). This suppression leads via the ISHE to a correction to the resistance of the NM layer. Similar to SMR, this new *Hanle magnetoresistance* (HMR) effect modulates the resistance of the NM layer with **H** (instead of **M**), exhibiting the same angular dependences as SMR: no resistance correction is observed for **H** parallel to **s**, whereas a resistance increase is obtained for **H** perpendicular to **s**. We analytically derive the equations that govern HMR allowing us to estimate the electronic diffusion coefficient in ultrathin films of Pt and *β*-phase Ta and to extract the spin diffusion length in *β*-Ta. The HMR effect, thus, opens a new avenue to study spin-to-charge current conversion in thin metal films with strong SOC.

All samples were prepared by patterning a Hall bar (width $W$=100μm and length $L$=800μm) on top of YIG [27] or nonmagnetic insulator (NMI) substrates (SiO$_2$, Pyrex or sapphire) via e-beam lithography (using PMMA), followed by sputter deposition of the metal (80W of power for Pt and 250W for Ta, and 3mtorr of Ar pressure, otherwise specified) with thickness $d_N$ and lift-off. Complementary data of all fabricated samples and control experiments can be found in the Supplemental Material [28]. Magnetotransport measurements were performed at different temperatures between 10 and 300K in a cryostat that allows applying magnetic fields up to H=9T and to rotate the sample 360º.

Figures 1(a)-1(d) show the longitudinal ($R_L$) and transverse ($R_T$) angular-dependent magnetoresistance (ADMR) curves in a Pt($d_N$=7nm)/YIG sample measured at 300K and two representative magnetic fields (1 and 9T) in the three relevant **H**-rotation planes. The rotation angles ($\alpha,\beta,\gamma$) and measurement configurations are defined in the sketches next to each panel. At these fields, **M** of YIG is saturated and follows the direction of **H** [28]. The ADMR measurements mostly follow the expected behavior described by the SMR theory [11,28]: (*i*) no ADMR is observed in $R_L(\gamma)$, (ii) a large modulation is observed in $R_L(\beta)$ and $R_L(\alpha)$, with the same amplitude $\Delta R_L$ and a $\cos^2(\alpha,\beta)$ dependence, and (iii) $R_T(\alpha)$ shows a $\sin(\alpha)\cos(\alpha)$ dependence, with an amplitude $\Delta R_T \sim \Delta R_L/8$, as expected from the geometrical factor $L/W \sim 8$. There is, however, one clear discrepancy with the SMR theory: the amplitude depends on the magnitude of the applied magnetic field, not expected for a saturated $\mathbf{M}$.

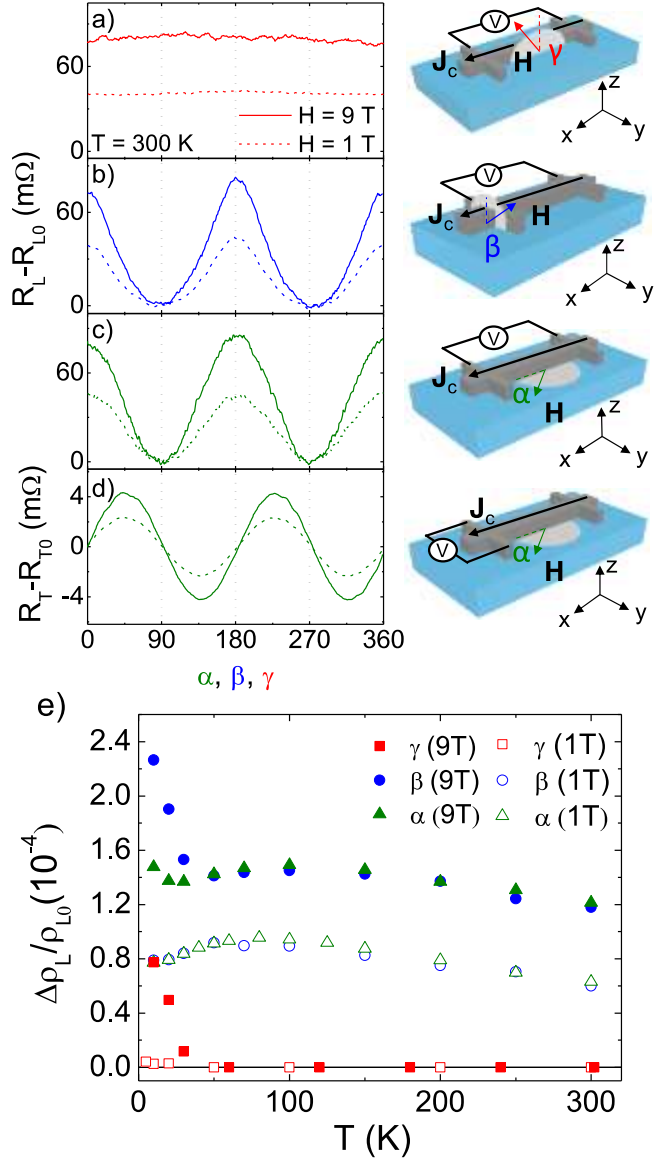

FIG. 1. (a)-(d) ADMR measurements in a Pt(7nm)/YIG sample at 300K and 1T (dashed lines) or 9T (solid lines) in the three relevant $\mathbf{H}$-rotation planes ($\alpha,\beta,\gamma$). Sketches on the right side indicate the definition of the angles, the axes and the measurement configuration. $R_{L0,T0}$ is the subtracted base resistance. (e) Temperature dependence of the longitudinal ADMR amplitude, normalized to the corresponding base resistance, obtained at 1 and 9T and for the three $\mathbf{H}$-rotation planes.

Further systematic ADMR measurements were performed between 10 and 300K (data obtained at 10 and 50K can be found in the Supplemental Material [28]). Figure 1(e) shows that the normalized ADMR amplitude $\Delta R_L/R_{L0} \equiv \Delta \rho_L/\rho_{L0}$ is identical for both $\alpha$– and $\beta$–rotation planes at 1T [28,38] and shows a small modulation with temperature –which can be attributed to the temperature evolution of the spin transport properties in Pt [8,39,40]–, whereas no ADMR is detected for $\gamma$, in agreement with SMR. Moreover, the larger MR at 9T is confirmed: a constant increase in the normalized ADMR amplitude is observed at all temperatures for both $\alpha$– and $\beta$–rotation planes. In addition to this, at $H$=9T, an extra ADMR contribution emerges below 50K, increasing as the temperature decreases, with the same amplitude for both $\beta$– and $\gamma$-rotation planes. This MR is due to the emergence of weak anti-localization (WAL) [28], an effect occurring in thin metal films with strong SOC at low temperatures [41–43].

In order to understand the increase in the ADMR amplitude between 1 and 9T, magnetic-field-dependent MR measurements were performed along the three main axes. Figure 2 shows normalized MR curves $[R_L(H_i)-R_{L0}]/R_{L0} \equiv \Delta\rho_L(H_i)/\rho_{L0}$, where $H_i$ denotes the magnetic field applied along the $i$ direction ($i=x,y,z$), obtained at 300K. Qualitatively similar curves were obtained at all temperatures between 50 and 300K. The dips and peak observed around zero magnetic field (see inset in Fig. 2) correspond to the magnetization reversal of the YIG substrate [28], in agreement with earlier reports [9,12–14]. At higher fields, $\Delta\rho_L(H_y)/\rho_{L0}$ keeps constant (zero) up to $|H_y|$=9T, but a parabolic-like increase is observed in both $\Delta\rho_L(H_x)/\rho_{L0}$ and $\Delta\rho_L(H_z)/\rho_{L0}$ for increasing $|H_{x,z}|$, an effect that cannot be explained via SMR. Similar magnetic field dependences of the MR at large fields have been already reported in Pt/YIG and attributed to MPE induced in the Pt layer, although no physical explanation of the observed directional dependence is given [22]. MPE is expected to be enhanced as the temperature decreases. However, the amplitude of our additional MR is fairly constant at all temperatures, excluding MPE as the origin of the effect.

Therefore, this additional MR effect must have another physical origin. Notice that in those directions where a MR is observed at high magnetic fields, **H** and **s** are perpendicular –**s** is parallel to the $y$ axis in our geometry–, whereas no resistance change is observed when **H** and **s** are collinear (see Fig. 2). This symmetry indicates that the high-field MR effect is originated by the interaction of the applied magnetic field with the generated spin current in the Pt layer *via* the Hanle effect [44]. This MR effect was originally predicted by Dyakonov [26] in 2007, although a different geometry was considered. A simple picture of this novel MR phenomenon is that the Hanle effect leads to a spin precession and dephasing of the spin accumulation generated *via* the SHE in our NM layer when **H** and **s** are not collinear. As a consequence, both the spin accumulation and the extra charge current produced *via* the ISHE are partially suppressed, thus, producing an increase of the resistance of the NM layer. Because of its origin, we call it *Hanle magnetoresistance*.

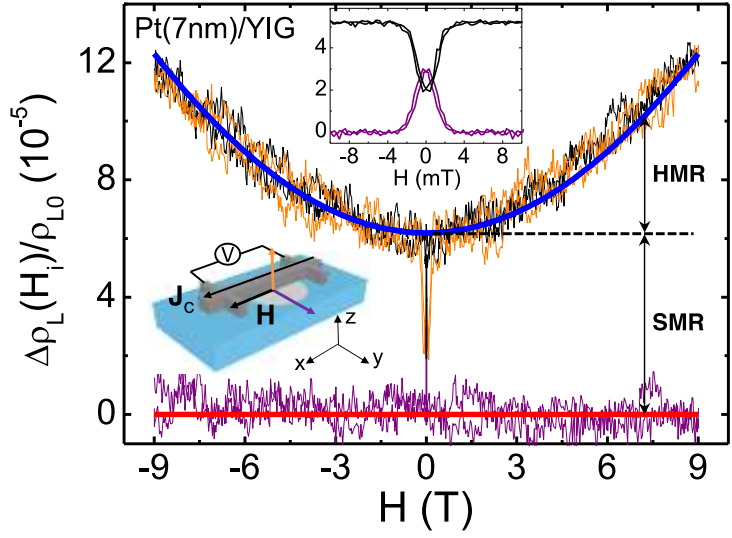

FIG. 2. Normalized magnetic-field-dependent MR measurements performed along the three main axes in Pt(7nm)/YIG at 300K (thin lines). See sketch for the definition of the axes, color code of the magnetic field direction and measurement configuration. Numerical computation of the MR curves by using Eq. (1) (parameters used are given in the text) are shown as thick red [$\Delta\rho_L(H_y)/\rho_{L0}$] and blue [$\Delta\rho_L(H_x)/\rho_{L0}=\Delta\rho_L(H_z)/\rho_{L0}$] lines. Both SMR and HMR contributions to the MR are schematically tagged. Inset: Zoom at low fields.

In order to confirm that HMR is at the origin of our experimental results, we derived the MR corrections starting from the kinetic equations for the charge and spin current densities in the presence of the spin Hall effect (quantified with the spin Hall angle $\theta_{SH}$) and a magnetic field **B** (**B**=$\mu$**H**, where $\mu \approx \mu_0$ is the magnetic permeability of Pt) [45]. The spin accumulation $\mu_{s,j}$ (where $j$ is the direction of spin polarization) depends only on $z$ and satisfies the diffusion equation [28,46]:

$$\partial_z^2 \mu_{s,j} - \frac{1}{\lambda^2}\mu_{s,j} + \frac{\omega_L}{D}\varepsilon_{jik}n_i\mu_{s,k} = 0, \qquad (1)$$

where $\lambda$ is the spin diffusion length, $D$ is the electron diffusion coefficient, $\omega_L=g\mu_B B/\hbar$ is the Larmor frequency (with $g$ the gyromagnetic g factor, $\mu_B$ the Bohr magneton and $\hbar$ the reduced Planck constant), $\varepsilon_{ijk}$ is the Levi-Civita tensor, and **n**=**B**/$B$ is a unit vector along the magnetic field. In the case of a NM/FMI bilayer, this equation can be solved with the following boundary conditions: i) At the boundary with the vacuum ($z=d_N$), the spin current vanishes and, ii) at the interface between the normal metal with the YIG ($z=0$), the spin density current is given in terms of the spin-mixing conductance $G_{\uparrow\downarrow}=G_r+iG_i$ [47]. We can numerically solve Eq. (1) to fit the MR curves shown in Fig. 2. For doing so, we first account for the SMR correction at small fields (B→0) and then fit the HMR contribution. Note that, in the approximation B→0, Eq. (1) leads to the standard SMR relation [11,28]. Using $\Delta\rho_L/\rho_{L0}$=6.16·10$^{-5}$, $\rho_{L0}$=6.31·10$^{-7}$Ωm, $d_N$=7nm, $\lambda$~1.3nm (considering that $\lambda \propto 1/\rho$ [48] and using values reported in Refs. [7,8,49]), $\theta_{SH}$=0.056 [49], and considering that $G_r$>>$G_i$ [50], Eq. (1) yields $G_r \approx 7.4 \cdot 10^{13} \Omega^{-1} m^{-2}$, which is within the range of reported values for Pt/YIG [9,10,12–14,51–54].

With the parameters given above, we can now fit the HMR contribution at high fields leaving $D$ as the free parameter [55]. The fitted curves, which nicely reproduce the observed $R_L(H_i)$ dependence, with no MR modulation for $R_L(H_y)$ and a quadratic MR correction for $R_L(H_x,H_z)$ (see Fig. 2), yield $D$=(6.4±0.6)·$10^{-6} m^2 s^{-1}$. Note that to calculate $D$ in Pt ultrathin films from the Einstein relation is not trivial since it is a two-band metal with a complex Fermi surface and, furthermore, the density of states can differ from the bulk value [56,57], which is the reason for the lack of reference values. Therefore, our model gives a powerful alternative to the current methods used to extract an effective diffusion coefficient in ultrathin metal films.

Importantly, in the absence of a FMI ($G_{\uparrow\downarrow}$=0 at z=0), the HMR effect will be also present at high fields. This is because HMR is intrinsic to the NM layer and does not require the presence of a FMI layer next to it. In this case, and at the leading order of $\theta_{SH}$<<1, Eq. (1) can be analytically solved. The longitudinal $\rho_L$ and transverse $\rho_T$ resistivities of the NM layer in the presence of $H$ read [28]

$$\rho_L = \frac{1}{\sigma_0} + \Delta\rho_0 + \Delta\rho_1(1 - n_y^2),$$
$$\rho_T = \Delta\rho_1 n_x n_y + \Delta\rho_2 n_z, \qquad (2)$$

with

$$\Delta\rho_0 = \frac{2\theta_{SH}^2}{\sigma_0} - \frac{2\theta_{SH}^2}{\sigma_0} \frac{\lambda}{d_N} \tanh\left(\frac{d_N}{2\lambda}\right),$$
$$\Delta\rho_1 = \frac{2\theta_{SH}^2}{\sigma_0} \left\{ \frac{\lambda}{d_N} \tanh\left(\frac{d_N}{2\lambda}\right) - \Re\left[\frac{\Lambda}{d_N} \tanh\left(\frac{d_N}{2\Lambda}\right)\right] \right\},$$
$$\Delta\rho_2 = \frac{2\theta_{SH}^2}{\sigma_0} \Im\left[\frac{\Lambda}{d_N} \tanh\left(\frac{d_N}{2\Lambda}\right)\right], \qquad (3)$$

where $\sigma_0 = 1/\rho_0 \approx 1/\rho_{L0}$ is the Drude conductivity, and $\frac{1}{\Lambda} = \sqrt{\frac{1}{\lambda^2} + i\frac{1}{\lambda_m^2}}$ with $\lambda_m = \sqrt{\frac{D\hbar}{g\mu_B B}}$. Note that Eq. (2) features the same angular dependence as the SMR equations [11]. The reason is that both effects have similar physical origins. They both rely on the interaction of the spin accumulation generated within the NM layer by the SHE with an external source. For SMR, the latter is given by $M$ of the FMI layer at the NM/FMI interface, whereas for HMR it is given by $H$, which acts in the entire NM layer. This explains, for instance, why the angular dependences in Figs. 1(a)-1(d) were preserved at large $H$. Note also that, to the first-order correction, the HMR amplitude $\Delta\rho_1$ goes as $(B/D)^2$ [28]. A discussion of the Hall-like term $\Delta\rho_2$ and its implications is presented in the Supplemental Material [28].

In order to confirm that HMR is an intrinsic feature of the NM layer, we grew Pt on NMI substrates and measured the MR at high fields. Figures 3(a) and 3(b) show the MR curves taken in Pt(7nm)/Pyrex and Pt(3nm)/Pyrex, respectively. These curves show the features predicted by our theory: (i) all $\Delta\rho_L(H_i)/\rho_{L0}$ curves start from zero at zero field, (ii) $\Delta\rho_L(H_y)/\rho_{L0}$ does not show MR, and (iii) both $\Delta\rho_L(H_x)/\rho_{L0}$ and $\Delta\rho_L(H_z)/\rho_{L0}$ show the same parabolic increase. Thick lines in Fig. 3(a) [3(b)] are fits of the experimental curves to Eqs. (2) and (3), obtaining $D=(6\pm1)\times10^{-6}$m$^2$s$^{-1}$ [$(3.4\pm0.5)\times10^{-6}$m$^2$s$^{-1}$] for $\rho_{L0}=8.97\cdot10^{-7}\Omega$m [$10.59\cdot10^{-7}\Omega$m], $d_N$=7nm [3nm], $\lambda$~0.9nm [~0.8nm] (values estimated from $\rho_{L0}$), $g\approx2$ [55], and $\theta_{SH}=0.056$ [49]. A summary of the results obtained for other Pt/NMI bilayers with varying Pt thickness, growth conditions and NMI substrate can be found in Table S1 of the Supplemental Material [28].

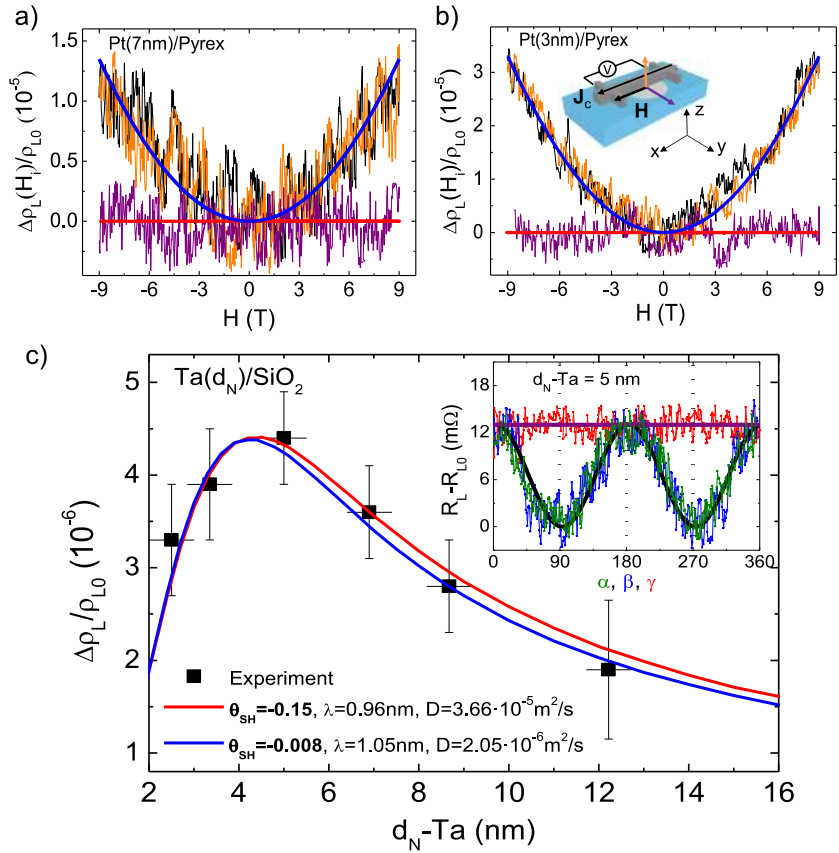

FIG. 3. (a), (b) Normalized magnetic-field-dependent MR curves in Pt(7nm)/Pyrex and in Pt(3nm)/Pyrex, respectively. Sketch in (b) defines the axes, color code of the magnetic field direction and measurement configuration. Fits of the MR curves to Eqs. (2) and (3) (parameters used are given in the text) are shown as thick red [$\Delta\rho_L(H_y)/\rho_{L0}$] and blue [$\Delta\rho_L(H_x)/\rho_{L0}=\Delta\rho_L(H_z)/\rho_{L0}$] lines. (c) HMR amplitude in Ta/SiO$_2$ as a function of Ta thickness at $H$=9T. Values are extracted via ADMR measurements. Red and blue lines

show fits of the experimental data to Eq. (3) for two different $\theta_{SH}$ values. Inset: ADMR measurements in Ta(5nm)/SiO$_2$ at 9T in α− (green line), β− (blue line) and γ− (red line) rotating planes. Fits of the curves to Eq. (2) are shown as thick black and purple lines. All measurements were taken at 100K to improve the signal-to-noise ratio.

A direct comparison of the HMR amplitude obtained in films growth on different substrates [see Figs. 2 and 3(a)] or under different deposition conditions is not straightforward because it yields different film roughness and/or average grain size (as confirmed via AFM and XRD measurements [28]), thus, modifying relevant microscopic parameters of the films [57]. However, we observed a consistent decrease in the extracted $D$ value for decreasing the grain size [28], which is one of the key parameters for the HMR effect to emerge. For instance, no HMR signal was detected in 7-nm-thick Pt films grown on SiO$_2$ using our standard conditions (see above), where the grain size was larger than in films grown on Pyrex or YIG, but a weak HMR signal was observed for films grown on SiO$_2$ at higher Ar pressures [28], which is known to promote smaller grains. It is also worth noting that we did not find a correlation between the measured $\rho$ and the extracted $D$ values. The reason they are decoupled in our ultrathin films is that, whereas the resistivity of the film is measured along the length of the Hall bar ($x$ direction), the relevant diffusion for the HMR effect is along the thickness of the film ($z$ direction). In ultrathin films, the diffusion in these two directions may begin to differ when the average grain size is on the order of the film thickness [28].

Finally, we further prove the robustness of the HMR effect by studying Ta thin films on SiO$_2$ [58]. Figure 3(c) shows the amplitude of the HMR effect ($\Delta R_L/R_{L0} \equiv \Delta \rho_L/\rho_{L0}$) as a function of Ta thickness extracted from ADMR measurements performed at $H$=9T. The inset shows the ADMR curves obtained in a 5-nm-thick Ta film. As expected from Eq. (2), the ADMR show a cos$^2$ dependence in both α− and β-rotation planes, while no modulation is observed for γ. The measured resistivity is fairly constant for all thicknesses, with a value $\rho_{L0} \sim 1.9 \times 10^{-6}\Omega$m and a weak temperature dependence [28], confirming our Ta films belong to the high-resistance β phase [59,60]. Since $\rho_{L0}$ does not change significantly with Ta thickness, we can fit the measured HMR amplitudes as a function of $d_N$ using the $\Delta \rho_1(d_N)$ dependence given in Eq. (3). For the fitting, we fix $\theta_{SH}$ and extract $\lambda$ and $D$. Fits using the largest ($\theta_{SH}$=-0.15 [61]) and the smallest ($\theta_{SH}$=-0.008 [62]) reported values in the literature, with $B$=9T and $g \approx 2$ [55], are shown in Fig. 3(c).

The nicely reproduced thickness dependence of the experimental MR amplitude by our model [see Fig. 3(c)] is additional strong proof of the existence of the HMR effect in thin metal films with strong SOC. Our fittings show that $\lambda$ can be extracted with good precision regardless of $\theta_{SH}$ because this parameter is basically constrained to the shape of the thickness dependence of the HMR effect, while $D$ (and $\theta_{SH}$) to its amplitude. Therefore, our fittings yield $\lambda$=(1.0±0.1)nm in our $\beta$-Ta films, a value consistent with the ones reported in the literature using other techniques [10,62,63] and suggest $D$ to be in the range $(2.1\times10^{-6}$-$3.7\times10^{-5})$m$^2$/s.

In conclusion, we report an effect predicted by Dyakonov [26], *Hanle magnetoresistance*, in thin metal films of Pt and $\beta$-Ta. The resistance of these films can be modulated with a magnetic field due to the suppression of the spin accumulation created by the spin Hall effect *via* the Hanle effect. This novel phenomenon provides a simple alternative way to quantify spin transport parameters such as the electronic diffusion coefficient, the spin Hall angle, or the spin diffusion length in ultrathin metal films with strong spin-orbit coupling without the need to involve ferromagnetic interfaces.

**ACKNOWLEDGEMENTS**

We thank Dr. L. Pietrobon and I. V. Tokatly for fruitful discussions. This work was supported by the European Union 7th Framework Programme under NMP Project No. 263104-HINTS and the European Research Council (Project No. 257654-SPINTROS), by the Spanish Ministry of Economy and Competitiveness under Project No. MAT2012-37638, FIS2011-28851-C02-02, No. FIS2014-55987-P, and No. MAT2013-46593-C6-4-P, and by the Basque Government under UPV/EHU Projects No. IT-756-13 and No. IT-621-13. M.I. and E.S. thank the Basque Government and the Spanish Ministry of Education, Culture and Sport, respectively, for a PhD fellowship (Grants No. BFI-2011-106 and FPU14/03102).

would simply yield to a normalization of the fitted *D* factor since they both enter together in Eqs. (2) and (3) as $D/g$.